\documentclass[useAMS,usenatbib]{mn2e}

\title[Reverse electron flow at polar caps]{Nuclear physics of reverse electron 
flow at pulsar polar caps}
\author[P. B. Jones]{P. B. Jones\thanks{E-mail:
p.jones1@physics.ox.ac.uk} \\
Department of Physics, University of Oxford, Denys Wilkinson Building,
Keble Road, Oxford, OX1 3RH, England}

\begin{document}

\date{}

\pagerange{\pageref{firstpage}--\pageref{lastpage}} \pubyear{}

\maketitle

\label{firstpage}

\begin{abstract}
Protons produced in electromagnetic showers formed by the reverse-electron flux are 
usually the largest component of the time-averaged polar-cap open magnetic-flux 
line current in neutron stars with positive corotational charge density. Although 
the electric-field boundary conditions in the corotating frame are 
time-independent, instabilities on both medium and short time-scales cause the 
current to alternate between states in which either protons or positrons and ions 
form the major component.  These properties are briefly discussed in relation to 
nulling and microstructure in radio pulsars, pair production in an outer gap, and 
neutron stars with high surface temperatures. 
\end{abstract}


\section{Introduction}

Polar-cap models of pulsar radio and X-ray emission all assume electron-positron 
pair creation by particles accelerated along the narrow bundle of open magnetic 
flux lines that intersect the light cylinder.  Understanding of the radio emission 
process remains limited. In a review of the subject, Melrose (1995) made the 
interesting observation that its characteristics are broadly unchanged over more 
than five orders of magnitude variation in magnetic dipole moment.  He also noted 
that whilst instabilities in the electron-positron plasma received much attention, 
there had been relatively less interest in the origin of the plasma.  But 
considerable progress with the latter problem has been made by Muslimov \& Tsygan 
(1992) who recognized that the Lense-Thirring effect, the general-relativistic 
dragging of inertial frames, has a profound influence on the acceleration electric 
field, which is the component ${\bf E}_{\parallel}$ locally parallel with ${\bf 
B}$, present in the frame of reference rotating with the neutron star.  The 
electric field is conservative in the rotating frame and is given in terms of the 
charge density $\rho$ by,
\begin{eqnarray}
\nabla\cdot{\bf E} = 4\pi\left(\rho - \rho_{0}\right),
\end{eqnarray}
where, in Euclidean space,
\begin{eqnarray}
\rho_{0} = - \frac{1}{4\pi c}\nabla\cdot\left(\left({\bf \Omega}\times{\bf 
r}\right)\times{\bf B}\right),    
\end{eqnarray}
is the density at which force-free corotation of the magnetosphere  exists 
(Goldreich \& Julian 1969).  The most significant way in which a 
general-relativistic treatment (see Muslimov \& Harding 1997; equations 37 and 38) 
modifies equation (2) is that the rotation angular velocity ${\bf \Omega}$ seen by 
a distant observer is replaced by ${\bf \Omega} - {\bf \Omega}_{\rm LT}$, where 
${\bf \Omega}_{\rm LT}$ is the Lense-Thirring angular velocity.  In the 
non-relativistic equation (1), both $\rho$ and $\rho_{0}$ have the same radial 
dependence because the angle between ${\bf \Omega}$ and ${\bf B}$ does not change 
much over distances no more than an order of magnitude greater than the polar cap 
radius.  But $\rho$ and $\rho_{0}$ have different physical origins, $\rho$ being 
determined principally by continuity, so that
the presence of the Lense-Thirring angular velocity
\begin{eqnarray}
{\bf \Omega}_{\rm LT} = \frac{2G}{c^{2}r^{3}}{\bf L},
\end{eqnarray}
where ${\bf L}$ is the angular momentum of the star, in the general-relativistic 
expression for $\rho_{0}$ 
produces a non-trivial radial dependence of the quantity $\rho - \rho_{0}$ which is 
the source of the acceleration field.

In the corotating frame, the potential boundary condition for equation (1) is $\Phi 
= 0$ on the neutron star surface and on the surface separating open from closed 
magnetic flux lines, with ${\bf E} = -\nabla\Phi$.  This is supplemented by the 
condition ${\bf E}_{\parallel} = 0$ on the neutron star surface if the work 
function there is not large enough to affect the free flow of charge.  Thus there 
are three possible cases: (i) ${\bf \Omega}\cdot{\bf B} > 0$, $\rho_{0} < 0$,  
${\bf E}_{\parallel} = 0$ with electron acceleration; (ii) ${\bf \Omega}\cdot{\bf 
B} < 0$, $\rho_{0} > 0$, ${\bf E}_{\parallel} = 0$, giving ion and positron 
acceleration  or (iii), ${\bf \Omega}\cdot{\bf B} < 0$ with ${\bf E}_{\parallel} 
\neq 0$.  The latter case is that considered originally by Ruderman \& Sutherland 
(1975) in which there is now renewed interest (see Gil et al 2008, and work cited 
therein) owing to the possibility that some neutron-star polar-cap magnetic fields 
may exceed the critical field $B_{c} = m^{2}c^{3}/e\hbar = 4.41\times 10^{13}$ G 
and may even
be one or more orders of magnitude larger than the dipole moment inferred from 
spin-down  indicates.  Recent calculations of ionic work functions (Medin \& Lai 
2006), including magnetic flux densities $B > 10^{14}$ G,
do not unequivocally exclude the neutron-star surface boundary condition ${\bf 
E}_{\parallel} \neq 0$.

Following the work of Muslimov \& Tsygan, case (i) appears consistent with electron 
acceleration and pair production occurring over a very wide range of surface 
magnetic flux densities.  Ion acceleration occurs in case (ii) with the possibility 
of positron production by secondary processes.  However, the time-independent 
boundary conditions $\Phi = 0$ and ${\bf E}_{\parallel} = 0$ applying in cases (i) 
and (ii) are very restrictive and in terms of electromagnetic considerations alone 
there is no obvious reason why either the current density or the nature of the 
electron-positron plasma within the open magnetic flux-line region should change so 
as to produce phenomena such as pulse nulling over times of the order of $10^{1} - 
10^{5}$s.  (For example, failure of the condition $\Phi = 0$ on the surface 
separating open from closed magnetic flux lines over such long time intervals 
appears unlikely.)
In this connection, it is worth noting that in PSR B1931+24, nulls last for 
intervals so long that it has been possible to measure the spin-down rate 
separately during the on and off-states of emission (Kramer et al 2006). Spin-down 
in the on-state is approximately twice as fast as in the off-states.  In each of 
cases (i) and (ii), the self-consistent current density in the open magnetic 
flux-line region (see Muslimov \& Harding 1997) does not deviate by so large a 
factor from the Goldreich-Julian value of $\rho_{0}c$.  Thus some change in the 
nature of the plasma is indicated.  Also there is the fact that the radio emission 
from many pulsars exhibits short-term variability in the form of both 
microstructure and subpulses. The microstructure may be caused by temporal 
variation of the accelerated plasma or may be a consequence of angular beaming, but 
there appears to be as yet no conscensus (see, for example, Smirnova et al 1994 and 
Lange et al 1998).
It is not obvious how any of these phenomena arise from electron or positron 
acceleration subject to time-independent boundary conditions.  For these reasons, 
the present paper addresses
the question of whether or not it is plausible that case (iii) actually exists in 
certain radio pulsars
and, specifically, to an investigation of nuclear processes in the electromagnetic 
showers produced by the reverse flow of electrons toward the polar cap.  In the 
course of this, we also find that case (ii) is actually less simple than might be 
supposed and that, even in the presence of time-independent boundary conditions, 
natural instabilities exist that cause medium and short time-scale variability in 
the nature of the accelerated plasma.

Nuclear processes were considered previously at magnetic flux densities $B \ll 
B_{c}$ in a less than satisfactory attempt to estimate backward photon production 
by the electromagnetic showers that are formed at the neutron-star surface (Jones 
1979).  But the present work is concerned with $B \sim B_{c}$ because only in this 
region is there a possibility that case (iii) could exist.
Also the problem here is much more simple: that of finding the rate of 
photo-production of protons by decay of the nuclear giant-dipole resonance (GDR), 
for which experimental data has greatly improved.  The mean electron-shower energy 
required to create a proton is a function of $B$ and of the nuclear charge
$Z$, but at $B > B_{c}$ is found to be no more than of the order of $2-5$ GeV.  
Diffusion to the neutron-star surface occurs with a short, but important, time 
delay and the process leads directly to an upper limit for a steady-state surface 
blackbody temperature in cases (ii) and (iii). But further analysis of the 
short-time instability in case (iii) shows that $T_{res}$, the polar-cap 
temperature in the absence of a reverse-electron flux, is actually the temperature 
which should be compared with the critical temperature $T_{c}$ inferred from 
calculated ionic work functions. 
The details of shower development and proton formation and diffusion leading to 
these conclusions are given in Sections 2 and 3.  The application to pulsar polar 
caps is described in Section 4.
 
Proton production by the reverse flux of electrons is prolific even at $B \ll 
B_{c}$ (see Section 2.5) and in Section 4.3 it is shown that it generally leads to 
medium time-scale instability in cases (ii) and (iii) producing similar variabilty 
in electron-positron pair formation.  Instability with a short time-scale is also 
likely to be a property of cases (ii) and (iii) but
unfortunately, is not possible to predict with complete confidence because it 
depends on details of the proton diffusion to the polar-cap surface that are not 
known with certainty.  However, it is of some interest in view of the observed 
characteristics of individual radio pulses in some pulsars.

The principal results of this paper are the proton formation rates at $B \sim 
B_{c}$ and the instabilities found in the fluxes of accelerated nuclei and protons 
and in pair formation.  Section 5 considers briefly the possibility that these may 
be connected with pulse nulling and microstructure, pair formation in any 
outer-magnetosphic gap, and the radio-observability of young neutron stars with 
high whole-surface temperatures.

\section[]{Electromagnetic shower development}

The later stages in the development of an electromagnetic shower whose primary 
particle of energy $E_{0} \sim 10^{2}-10^{3}$ GeV moves parallel with ${\bf B}$ are 
determined principally by Coulomb pair production, Compton scattering, and by 
magnetic bremsstrahlung activated by multiple Coulomb scattering.  Formation of the 
giant-dipole state is by far the most important nuclear interaction and occurs 
predominantly in the very late stages of shower development, within the photon 
energy band of $15 - 30$ MeV.  Higher-multipole interactions have much smaller 
cross-sections and appear only at greater energies where the photon track length 
within the shower is smaller.  The principal processes will be described in some 
detail in Sections 2.1 - 2.4, but in Section 2.4, other processes that are present 
in showers at $B > B_{c}$ are discussed briefly.

The natural unit of length for shower development at $B \ll B_{c}$ is the radiation 
length defined in terms of the Bethe-Heitler formulae (Bethe 1934) for the 
bremsstrahlung and pair production cross-sections.  This convention is adopted here 
for convenience, though with cross-sections modified by a screening wavenumber 
defined by the matter density at the neutron-star surface.  For $10 < B_{12} < 
10^{3}$, where $B_{12}$ is the magnetic flux density in units of $10^{12}$ G, we 
fit the ion number densities found by Medin \& Lai (2006) by the expression
\begin{eqnarray}
N = 2.7\times 10^{25}Z^{-0.7}_{26}B^{1.2}_{12} {\rm cm}^{-3}.
\end{eqnarray}
The radiation length is then,
\begin{eqnarray}
l_{r} = 2.4\times 10^{-2}Z^{-1.3}_{26}B^{-1.2}_{12}
\left(\ln\left(62Z^{1/2}_{26}B^{-1/2}_{12}\right)\right)^{-1}{\rm cm}
\end{eqnarray}
in which the logarithm, whose argument is derived from the infinite linear-chain 
ion separations given by Medin \& Lai, replaces the $\ln(183Z^{-1/3})$ of the 
Bethe-Heitler formulae.
This is extremely small; $l_{r} = 5.2 \times 10^{-5}$ cm at $B = 10^{14}$ G and $Z 
= 26$.
The Bethe-Heitler mean free path for Coulomb pair production is $9l_{r}/7$.

The electron or positron dispersion relation in a field of magnetic flux density 
${\bf B}$ is
\begin{eqnarray}
E = \left(1 + p_{\parallel}^{2} + 2n\frac{B}{B_{c}}\right)^{1/2}.
\end{eqnarray}
Unless otherwise stated,  all electron or photon energies and momenta will be 
expressed in units of the electron rest energy and of $mc$.
The classical momentum components $p_{\parallel}$ and $p_{\perp}$ are parallel with 
and perpendicular to ${\bf B}$, and $n$ is the Landau quantum number.  The 
conserved quantities in electron-photon interactions are the total energy and the 
parallel momentum component, but not $p_{\perp}$ except in the correspondence 
principle limit of $n \gg 1$.   States of $n \sim 1$ are important at $B > B_{c}$, 
and are best regarded as of one-dimensional motion, given by $p_{\parallel}$, with 
effective mass $\sqrt{1 + 2nB/B_{c}}$.  It is unfortunate that there appear to be 
no published calculations of the Coulomb bremsstrahlung and pair creation cross 
sections that are valid for $B \sim B_{c}$.

\subsection{Magnetic bremsstrahlung}

Multiple Coulomb scattering of a shower electron or positron is a sequence of 
ascending (or descending) Landau transitions from its creation state $n_{\pm}$.
For $B > B_{c}$, the classical $(n\gg 1)$ energy loss-rate expression for magnetic 
bremsstrahlung (see the reviews of Erber 1966 and of Harding \& Lai 2006) is not 
valid, particularly in the later stages of shower development where the dominant 
process has two stages: Coulomb excitation of the $n = 0\rightarrow 1$ transition 
followed by cyclotron emission. 
In order to see that this is so, and in the absence of an accurate expression, we 
are obliged to adopt the correspondence principle for an order of magnitude 
estimate and therefore assume that a momentum transfer $q_{\perp} \sim 
\sqrt{2B/B_{c}}$ from the nuclear Coulomb field is necessary for an $n = 
0\rightarrow 1$ transition.  The mean free path, derived from the Rutherford 
scattering formula for $q_{\perp}$ and from the number density of equation (4), is 
then, 
\begin{eqnarray}
\lambda = 5.1\times 10^{-5}B_{12}\ln \left(62Z^{1/2}_{26}B^{-1/2}_{12}\right)l_{r},
\end{eqnarray}
equivalent to $\sim 10^{2}$ Coulomb excitations per radiation length at $10^{14}$ 
G.  The relativistic quantum-cyclotron emission rate can be approximated by 
$\tau^{-1}_{ce} \approx 1.0\times 10^{-3}\omega_{B}$ in the $p_{\parallel} = 0$ 
frame of reference, where $\omega_{B}$ is the classical cyclotron angular 
frequency.
This expression is almost stationary with respect to $n$ and $B/B_{c}$ in the 
vicinity of $10^{14}$ G (see Fig. 3 of Harding \& Lai 2006).  The condition that 
the cyclotron emission rate exceeds the Coulomb excitation rate is that,
\begin{eqnarray}
\gamma_{0} < \frac{\lambda}{c\tau_{ce}} \approx 0.7B^{0.8}_{12}Z^{-1.3}_{26},
\end{eqnarray}
in which $\gamma_{0}$ is the Lorentz factor for transformation from the 
$p_{\parallel} = 0$ frame to that of the rotating star. Thus the statement that $n 
= 0 \rightleftharpoons n = 1$ transitions are the dominant processes is valid in 
the very late stages of shower development at $10^{14}$ G and we can conclude that 
in this region electrons or positrons lose their energy over distances one or two 
orders of magnitude smaller than $l_{r}$.

From the kinematic conservation laws,
the maximum photon transverse momentum (that is, $\perp$ to ${\bf B}$) in cyclotron 
emission for an $n \rightarrow n^{\prime}$ transition is,
\begin{eqnarray}
k_{\perp} = \sqrt{1 + 2n\frac{B}{B_{c}}}
      - \sqrt{1 + 2n^{\prime}\frac{B}{B_{c}}}.
\end{eqnarray}
Because the low-energy part of the electron-positron spectrum is the source of most 
GDR-band photons, we shall show, in Section 2.3, that this limit on photon 
transverse momentum is of crucial importance in estimating their total track length 
in the shower.  But earlier in the shower, for values of $\gamma_{0}$ about two 
orders of magnitude larger such that $\gamma_{0}c\tau_{ce} > l_{r}$, Coulomb 
bremsstrahlung is the more important energy-loss process.

\subsection{Coulomb bremsstrahlung}

The cross-section for Coulomb bremsstrahlung at $B > B_{c}$ must differ from the 
Bethe-Heitler formula, but no result appears to have been published at the present 
time. For the emission of a photon of momentum ${\bf k}$ at an angle $\theta$ with 
${\bf B}$, by an electron or positron of initial momentum $p_{\parallel}$ with a 
change of Landau quantum number $n \rightarrow n^{\prime}$,  the required momentum 
transfer components are, in the limit $p_{\parallel} \gg 1$,
\begin{eqnarray}
q_{\parallel}  & = &  k(1 - \cos\theta) + (n^{\prime} - 
n)\frac{B}{B_{c}}\frac{1}{p_{\parallel}} +   \nonumber     \\ 
	&  &  \frac{k}{2p^{2}_{\parallel}}\left(1 + 2n^{\prime}\frac{B}{B_{c}}\right),   
\nonumber  \\
q_{\perp} & \sim  & \sqrt{\frac{2B}{B_{c}}}, \hspace{1cm} {\rm for} \hspace{5mm} 
n^{\prime} \neq n
\end{eqnarray}
For transitions with $n^{\prime} \neq n$, the threshold longitudinal momentum 
transfer is larger than its zero-field value of $q_{\parallel} = 
k/2p^{2}_{\parallel}$ and it is therefore possible that the true cross-section is 
reduced in comparison with the Bethe-Heitler value.  But this merely increases the 
electron track length in the earlier stages of shower development, with which we 
are not concerned, and does not affect the photon total track length in the GDR 
band.

\subsection{Coulomb and magnetic pair creation}

A photon with momentum ${\bf k}$ at an angle $\theta$ with ${\bf B}$ can convert to 
an electron-positron pair with Landau quantum numbers $n_{\pm}$.   Only the 
longitudinal component of momentum transfer from the nuclear Coulomb field appears 
in the kinematic conservation laws and the threshold is,
\begin{eqnarray}
q_{\parallel} & = & k(1 - \cos\theta) - \frac{1}{2p_{\parallel +}}\left(1 + 
  2n_{+}\frac{B}{B_{c}}\right) -   \nonumber  \\
              &   &      \frac{1}{2p_{\parallel -}}\left(1 + 
  2n_{-}\frac{B}{B_{c}}\right),
\end{eqnarray}
which is valid in the limit $p_{\parallel\pm} \gg 1$ and has two regions of 
interest; $q_{\parallel} = 0$ and $q_{\parallel} \neq 0$.  It is again unfortunate 
that there appears to be no published cross-section for the Coulomb process at 
finite $q_{\parallel}$, and in finding the total photon track length in this case, 
we shall assume the Bethe-Heitler  mean free path, multiplied by an unknown 
correction factor $\eta_{p} > 1$.  However, $q_{\parallel} = 0$ is possible  at a 
finite $\theta$, and the associated transverse momentum component $k_{\perp}$ is 
the threshold for single-photon magnetic pair production. The individual thresholds 
that are important here are,
\begin{eqnarray}
k_{\perp} &  =  &  2,   \hspace{3cm} n_{+} = n_{-} = 0,  \nonumber  \\
k_{\perp} &  =  &  1 + \sqrt{1 + 2\frac{B}{B_{c}}}, \hspace{10mm} n_{\pm} = 0, 
\hspace{2mm}n_{\mp} = 1.
\end{eqnarray}
Transitions to the Landau state $n_{-} = n_{+} = 0$ are not allowed for photons 
polarized so that the electric vector is parallel with ${\bf k}\times{\bf B}$ 
(Semionova \& Leahy 2001, see also Harding \& Lai 2006), but are allowed for 
photons polarized with electric vector perpendicular to ${\bf k}\times{\bf B}$.  
The classical expression  for the conversion rate, which is valid in the $n_{\pm} 
\gg 1$ limit, is a function of $k_{\perp}B/B_{c}$ (see Erber 1966; Harding \& Lai 
2006), and is not applicable here at fields $B > B_{c}$ with 
$n_{\pm} \sim 1$.  But the rate $R_{\gamma ee} = 1.2\times 10^{6}$ cm$^{-1}$ that 
it predicts for $30$ MeV photons at $10^{14}$ G, immediately above the higher of 
the two thresholds given by equations (12), is so large that, whatever its 
inadequacies, it must be presumed that magnetic conversion occurs within a very 
small fraction of a radiation length.  Thus photons with $k_{\perp}$
above the higher threshold do not contribute significantly to the total photon 
track length or to the photoproduction of protons through GDR formation.  Photons 
in the allowed polarization state can convert above the lower threshold but the 
rate is reduced by a factor of the order of $(k_{\perp}/k)^{2} \approx 10^{-3}$ and 
is approximately an order of magnitude smaller than that for Coulomb conversion.  A 
further small contribution to the magnetic conversion rate must exist because the 
correct electron and positron propagators are those for passage through condensed 
matter, not free space.  Therefore, Coulomb interaction introduces a small 
component of $n_{\pm} = 0$, $n_{\mp} = 1$ into the $n_{+} = n_{-} = 0$ state.  But 
this is relatively unimportant and has not been studied in detail here.

Comparison of equations (9) and (12) shows that all photons emitted by cyclotron 
decay of Landau states $n$ to the ground state $n = 0$ do not exceed the thresholds 
for subsequent single-photon magnetic pair production provided,
\begin{eqnarray}
\sqrt{1 + 2n\frac{B}{B_{c}}} & <  &  3,     \nonumber   \\
\sqrt{1 + 2n\frac{B}{B_{c}}} & <  &  2 + \sqrt{1 + \frac{2B}{B_{c}}},
\end{eqnarray}
where the two inequalities refer to the two thresholds of equation (12).  The 
angular distribution of the cyclotron decay photon in the $p_{\parallel} = 0$ frame 
of reference considered in Section 2.1 is approximately isotropic (see Fig. 3 of 
Latal 1986), so that because equation (9) gives the maximum possible  value of 
$k_{\perp}$, photons with $k_{\perp}$ below the maximum may well fall below the 
thresholds (12)
even if the above inequalities are not satisfied.  It is interesting to note the 
first inequality fails for the minimum Landau quantum number $n = 1$ for all $B > 
4B_{c}$ whilst the second is satisfied for all $B$.

\subsection{Compton scattering and other processes}

Compton scattering is the most important of the processes of second order in the 
fine structure constant that are not considered in previous Sections.  It can 
significantly modify the very late stages of shower development at $B > B_{c}$.
Extensive cross-section calculations have been performed by Gonthier et al (2000) 
for a wide interval of photon energy.  For the purposes of estimating its 
significance, the mean free path equivalent to the Thomson cross-section 
$\sigma_{T}$ for the electron density given by equation (4) with $B = 10^{14}$ G 
and $Z = 26$ is $\lambda_{T} = 8.5 \times 10^{-6}$ cm, which is less than a 
radiation length.  The total cross-sections (see Fig. 2 of Gonthier et al) are 
negligible at photon energies $k > 10^{3}$, but are significant in the GDR band.  
Scattered photons with $k_{\perp}$ exceeding the higher threshold (12) undergo 
prompt magnetic conversion.  Gonthier et al have shown that the exact cross-section 
in this region is well approximated by the non-magnetic Klein-Nishina formula, a 
very convenient result which enables us to establish that scattered photons with 
$k_{\perp}$ below the magnetic pair thresholds comprise only a very small, and for 
present purposes, negligible fraction of the total Compton cross section.  Thus the 
cross-sections that effectively convert photons to electron-positron pairs can be 
obtained directly from Fig. 2 of Gonthier et al.  They are slowly varying functions 
of $B/B_{c}$.  We shall assume a photon energy of $21$ MeV for GDR formation in 
nuclei of atomic number $10 \leq Z \leq 26$ (see the review of Hayward 1963).  At 
this energy, the Compton cross sections are $\sigma_{C} = 3.6\times 
10^{-2}\sigma_{T}$ for $B = B_{c}$ and $4.8\times 10^{-2}\sigma_{T}$ for $B = 
10B_{c}$.  At higher energies, the cross section decreases, being approximately 
$\propto k^{-1}$.

The other second-order processes which should be considered briefly involve the 
blackbody radiation field, whose photons are typically of energy $k_{bb}\sim 
10^{-3}$. Inverse Compton scattering can give photons exceeding the pair production 
thresholds (12), but only for electron momenta greater than $k\sim 10^{4}$.  
Similarly, two-photon pair production is possible, but with a threshold $k = 
k_{bb}^{-1}$ and a negligible transition rate because the blackbody photon number 
density is many orders of magnitude smaller than the ion densities given by 
equation (4).  Direct production of electron-positron pairs (the trident process) 
is of third order and has no significant effect on shower development.  Photon 
splitting, also of third order, has a small transition rate, saturating at 
$R_{ps}\approx 0.4k^{-1}k_{\perp}^{6}$ cm$^{-1}$ at $B > B_{c}$ (see the review of 
Harding \& Lai 2006).

The remaining process that does change shower development is the 
Landau-Pomeranchuk-Migdal effect for which we refer to the extensive review given 
by Klein (1999), also to the work of Hansen et al (2004) for its laboratory 
verification at electron energies of several hundred GeV.  In condensed matter at 
zero field, multiple Coulomb scattering removes the coherence that the 
Bethe-Heitler formulae assume over distances of the order of $q_{\parallel}^{-1}$, 
where $q_{\parallel}$ is the very small longitudinal momentum transfer from the 
nucleus, and so reduces the cross sections, particularly for bremsstrahlung 
production of low-energy photons.
This is almost certain to be a feature of the showers considered here owing to the 
high density of matter at the neutron-star surface. But it is not possible to make 
even qualitative estimates of the effect because the nature of multiple Coulomb 
scattering is changed at $B > B_{c}$ (see Section 2.1) and there is also a strong 
one-dimensional ordering of ions parallel with ${\bf B}$ arising from the formation 
of bound moleculsr chains (see Medin \& Lai 2006) which will be of significance for 
questions of coherence of the Coulomb bremsstrahlung and pair creation  amplitudes.  
It is almost certain that in the early stage of shower development considered here, 
the LPM effect leads to an inward displacement of the GDR photon band region, 
though without significant change in its total track length.  We shall see in 
Section 3 that the formation of protons is also not significantly changed but that 
their time-scale for diffusion to the surface is modified in a non-trivial way as 
described in Section 4.3.

\subsection{Photon track length in a shower}

The distribution of photon track length is found from electromagnetic shower 
theory.  On dimensional grounds, the track length per unit interval of photon 
energy $k$ such that $E_{c} \ll k \ll E_{0}$ in a shower of primary electron energy 
$E_{0}$ must be of the form,
\begin{eqnarray} 
G(k) = \frac{yE_{0}l_{r}}{k^{2}}
\end{eqnarray}
in which $y$ is a dimensionless parameter and $E_{c}$ is the critical energy at 
which electron or positron ionization and bremsstrahlung energy-loss rates are 
equal.  In the zero-field case, where there is more certainty in the fundamental 
processes of shower development, it is possible to obtain reliable values of $y$.  
But at $B > B_{c}$,  we are obliged to adopt a more elementary and analytical 
approach, usually referred to as Approximation A, which includes only the most 
significant processes which are bremsstrahlung and pair creation (Landau \& Rumer 
1938).
But even at this level, there are difficulties because the relative importance of 
Coulomb and magnetic bremsstrahlung changes as the shower develops.  The later 
stages of shower development with which we are concerned here are almost completely 
determined by magnetic bremsstrahlung, as described in Section 2.1.
Therefore,
a modified Approximation A, in which throughout the development of the shower, 
Coulomb bremsstrahlung is replaced by magnetic bremsstrahlung, has been used to 
calculate an approximate value of $y$.  Pair creation is assumed to be determined 
by the Bethe-Heitler formula with the unknown correction factor $\eta_{p}$ retained 
(Section 2.3). To obtain the total photon track length, only the form of the photon 
spectrum is needed and initially,
we assume the number density $\propto k^{-2/3}$ per unit interval of $k$ which is 
valid in the classical magnetic bremsstrahlung limit of $n \gg 1$.  Following 
closely the procedures of Landau \& Rumer (1938) and Nordheim \& Hebb (1939), we 
find that $G(k)$ is given by the inverse Mellin transform,
\begin{eqnarray}
G(k) = \frac{-1}{2\pi i}\int^{i\infty + \zeta}_{-i\infty 
+\zeta}\frac{\eta_{p}l_{r}E^{s}_{0}}{k^{s+1}D(s)}ds,
\end{eqnarray}
in which
\begin{eqnarray}
D(s) & = &  \left(\frac{2}{s+1} - \frac{8}{3}\frac{1}{s+2}
+ \frac{8}{3}\frac{1}{s+3}\right) \nonumber \\
     &   &+ \frac{7}{9}\left(s+\frac{1}{3}\right)
\int^{1}_{0}\frac{\xi^{s}-1}{(1-\xi)^{2/3}}d\xi
\end{eqnarray}
with $\zeta$ chosen so that all poles lie to the left of the contour.  To obtain 
the term linear in $E_{0}$ we retain only the contribution from the $s=1$ pole and 
find $y=0.608\eta_{p}$. Apart from the unknown factor, this differs little from the 
zero-field Approximation A in which, for an approximate photon number density 
$\propto k^{-1}$, the value is $y = 0.47$.  

But as shown in Section 2.1, the magnetic bremsstrahlung classical limit is not 
reached in the later stages of shower development. Instead, the dominant processes 
are Landau transitions $n = 0 \rightleftharpoons n = 1$ in which the angular 
distribution of the cyclotron decay photons in the $p_{\parallel} = 0$ frame of 
reference is approximately isotropic. Thus a Lorentz transformation to the frame of 
the rotating star produces a photon number density distribution which is a constant 
per unit interval of $k$.  In this case, the denominator function $D(s)$ in 
equation (15) is replaced by
\begin{eqnarray}
D_{c}(s) = \frac{2}{s+1} - \frac{8}{3}\frac{1}{s+2} + \frac{8}{3}\frac{1}{s+3}
- \frac{7}{9}s,
\end{eqnarray}
giving the result $y = 0.871\eta_{p}$.
It is worth noting that the Approximation A results for total photon track length 
are independent of the absolute rates of bremsstrahlung emission.  Even the effects 
of quite substantial changes in its spectral shape are relatively small, being 
limited to small increases in $y$ for the harder spectra.

Two processes not included in Approximation A, Compton scattering and energy loss 
by ionization, can affect the late stages of shower development. 
The principal effect of Compton scattering is a reduced effective mean free path 
for pair creation by photons in the GDR band and therefore, a decreased total 
photon track length available for GDR formation.  This is important and can be 
allowed for by the replacement,
\begin{eqnarray}
\eta_{p} \rightarrow \frac{\eta_{p}}{1 + \eta_{p}x_{C}}, \hspace{1cm} x_{C} = 
\frac{9}{7}Nl_{r}Z\sigma_{C},
\end{eqnarray}
in equation (15), with $\sigma_{C}(k)$ evaluated at the GDR formation energy $k_
{gd} = 41$.

It is known through comparison of Approximations A and B that in the zero-field 
case, electron and positron ionization energy loss produces some reduction in $y$ 
because GDR-band energies are of the same order of magnitude as the critical energy 
$E_{c}$.  But the total electron-positron track length in the later stages of 
shower development at $B > B_{c}$ is much reduced because energy loss by magnetic 
bremsstrahlung predominates and occurs over lengths one or two orders of magnitude 
smaller than $l_{r}$ (see Section 2.1). Thus the relative importance of ionization 
energy loss is much reduced near the GDR band and we attempt no correction for it.

On this basis, we shall adopt the track-length estimate given by equation (14) in 
which $yl_{r}$ is found from equations (4), (5), (15), (17) and (18), with the 
parameter $\eta_{p}$ as an unknown.

\section[]{Proton production}

The mean number of protons formed in a shower per unit interval of primary electron 
energy is
\begin{eqnarray}
W_{p} = \langle NI_{\gamma p}\rangle G(k_{gd})/E_{0},
\end{eqnarray}
in which $I_{\gamma p}$ is the energy-integrated cross section for giant dipole 
resonance formation and decay into channels containing one or more protons, and
the square brackets denote an average over the depth distribution of GDR-band 
photons in the shower.  This latter function is unknown in detail, owing to the 
gradual transition from Coulomb to magnetic bremsstrahlung as the dominant process 
in shower development and to
 the existence both of the unknown parameter $\eta_{p}$ and the
  LPM effect, but it is certainly very small near the surface and increases to a 
maximum at a depth which is likely to be of the order of $10l_{r}$.  Therefore, we 
are obliged to evaluate the ion number density $N$ and $I_{\gamma p}$ at a mean 
atomic number $Z$.  The integrated cross-section is expressed as the 
electric-dipole sum-rule value (Hayward 1963) multiplied by a factor $x_{p}$,
\begin{eqnarray} 
I_{\gamma p} = 117.4\left(x_{p} Z\right)\left(1 - \frac{Z}{A}\right)\hspace{3mm} 
mc^{2}\hspace{3mm} {\rm  mb}
\end{eqnarray}
where $A$ is the nuclear mass number. Measured cross-sections for light elements 
$10 \leq Z \leq 26$ provide evidence for a value $x_{p}\approx 0.5$ (Hayward 1963; 
Kerkhove et al 1985; Assafiri \& Thompson 1986). But in the circumstances of shower 
development below the neutron-star surface, the problem of estimating the effective 
value of $x_{p}$ is more complicated. 

The protons formed in GDR decay lose energy by ionization, are quickly thermalized, 
and diffuse to the surface with negligible probability of secondary nuclear 
interaction.  However, this is not so in the case of GDR-decay neutrons which lose 
energy and diffuse to the surface by a sequence of nuclear elastic scatterings in 
the course of which capture in a transition $A \rightarrow A + 1$ may occur with a 
significant probability which is, however, hard to estimate.  Consequently, nuclei 
in the shower region may be neutron-rich.  There is some evidence (Fultz et al 
1974; O'Keefe et al 1987) that $x_{p}$ decreases with increasing nuclear asymmetry 
$A - 2Z$, possibly owing to the associated divergence between neutron and proton 
separation energies.  A simple model of the effect of this increase in nuclear 
asymmetry is as follows.  The energy-integrated cross-section for neutrons is given 
by equation (20) with the substitution $p\rightarrow n$ and $x_{p} + x_{n} = a 
\approx 1$.  Suppose that neutrons diffuse to the surface with survival probability 
$\epsilon_{n}$ and that the captured fraction $1 - \epsilon_{n}$ increases nuclear 
asymmetry.  This produces a bias toward GDR decay by neutron emission.  To take 
account of this,
let $x_{n} = Hx_{p}$, in which $H$ is a function of $A - 2Z$.  The steady-state 
condition for equal neutron and proton loss rates is then $x_{p} = 
\epsilon_{n}x_{n} = H\epsilon_{n}x_{p}$, from which we find that $\epsilon_{n}$ 
determines the asymmetry function $H$ and $x_{p} = a\epsilon_{n}(1 + 
\epsilon_{n})^{-1}$.  In the limit of small $\epsilon_{n}$, the nuclear asymmetry 
is, of course, limited by $\beta$-instability, the relevant time-scale being  that 
in which the corotational current density $\rho_{0}c$ removes the equivalent of one 
radiation length of matter, $t_{rl} = Nl_{r}AeP/B \sim 10^{2}$ s for $B = 10^{14}$ 
G and rotation period $P = 1$ s.

Values of $W_{p}$ found from equation (19) are given in the final column of Table 1 
for intervals of $Z$ and of $B$ that might permit the ${\bf E}_{\parallel} \neq 0$ 
boundary condition at the polar cap. They assume the parameters $\eta_{p} = 1$ and 
$x_{p} = 0.5$.  To allow for the effect of neutron capture, the numbers in the 
right-hand column should be multiplied by $2a\epsilon_{n}(1 + \epsilon_{n})^{-1}$.  
Multiplication by $\eta_{p}(1+x_{C})(1+\eta_{p}x_{C})^{-1}$ allows for values of 
$\eta_{p} > 1$.  In the latter case, it is worth noting that even in the limit
$\eta_{p} \gg 1$, the existence of the Compton effect, which has been extensively 
studied at $B > B_{c}$, maintains shower development.
Proton formation rates in the Table do not vary greatly as functions of $Z$ or of 
$B$ and are extremely high, so that a primary electron of $10^{3}$ GeV creates 
between $200$ and $500$ protons. It is, of course, the case that many 
approximations have been made in Sections 2.1 - 2.4 but these do not have a great 
effect on the late stages of shower development in which most of the total track 
length given by equation (14) is generated.  It is principally for this reason that 
we believe Table 1 provides a reliable estimate of the true rates.

\begin{table}

\begin{minipage}{8cm}
\caption{The number of protons created per unit primary electron energy, in units 
of $mc^{2}$, are given in the right-hand column for values of the mean nuclear 
charge $Z$ in the region of the shower maximum and of the magnetic flux density $B$ 
in units of the critical field $B_{c} = 4.41\times 10^{14}$ G.  Also given are the 
radiation length $l_{r}$ in neutron-star surface matter and the parameter $x_{C}$ 
that allows for the Compton effect correction given by equation (18). It is assumed 
here that the correction factor applied to the Bethe-Heitler pair creation mean 
free path is $\eta_{p} = 1$ and that the strength of the giant dipole-resonance 
proton decay channels is given by $x_{p} = 0.5$.  We refer to Section 3 for the 
correction to be applied given different values of these two parameters.  }

\begin{tabular}{@{}lrrrr@{}}
\hline
  $Z$  &  $BB^{-1}_{c}$  &  $l_{r}$  &  $x_{c}$  &  $W_{p}$  \\
	 &               &  ${\rm cm}$ &       &  $(mc^{2})^{-1}$   \\
\hline
  10   &   1   &  $5.0\times 10^{-4}$    &  0.76    &  $2.1\times 10^{-4}$   \\        
  10   &   3   &  $2.0\times 10^{-4}$    &  1.30    &  $2.4\times 10^{-4}$  \\
  10   &   10  &  $9.2\times 10^{-5}$    &  2.96    &  $2.8\times 10^{-4}$  \\
  18   &   1   &  $2.0\times 10^{-4}$    &  0.37    &  $1.5\times 10^{-4}$  \\
  18   &   3   &  $7.3\times 10^{-5}$    &  0.58    &  $1.7\times 10^{-4}$  \\
  18   &   10  &  $2.9\times 10^{-5}$    &  1.11    &  $2.2\times 10^{-4}$  \\
  26   &   1   &  $1.1\times 10^{-4}$    &  0.23    &  $1.0\times 10^{-4}$  \\
  26   &   3   &  $4.0\times 10^{-5}$    &  0.36    &  $1.2\times 10^{-4}$  \\
  26   &   10  &  $1.5\times 10^{-5}$    &  0.64    &  $1.5\times 10^{-4}$  \\
\hline
\end{tabular}
\end{minipage} 
\end{table}

\subsection{Proton diffusion}

Diffusion of protons to the surface is very rapid.  The distribution of proton
formation depth depends on a number of factors and is not easily obtained from 
shower calculations such as those described in Section 2.5. The dominant process of 
photon creation changes during shower development from Coulomb to magnetic 
bremsstrahlung (see Sections 2.1 and 2.2).  The LPM effect, considerd in Section 
2.4, displaces shower development inward as does the unknown factor $\eta_{p} > 1$, 
particularly the region of low-energy photon formation which is relatively compact. 
On this basis, it is convenient to assume a normalized proton formation depth 
distribution given by
\begin{eqnarray}
g_{p}(z) = \delta(z - z_{p}),
\end{eqnarray}
though with the reservation that $z_{p}$ is not well known. Let the time interval 
between creation and arrival at the surface be $t - t^{\prime} = \tau t_{p}$ with 
$\tau$ dimensionless.  For a semi-infinite  homogeneous medium, the probability per 
unit interval of $\tau$ that a proton created at depth $z < 0$ arrives at the 
surface $z = 0$ is,
\begin{eqnarray}
P(\tau,z) = -z\left(\frac{2}{3}\pi\tau^{3}z^{2}_{p}\right)^{-1/2}
\exp\left(-\frac{3z^{2}}{2z^{2}_{p}\tau}\right),
\end{eqnarray}
in which $z^{2}_{p} = 3Dt_{p}$ and $D$ is the diffusion coefficient.    Thus the 
probability of arrival at the surface with delay $\tau$ is,
\begin{eqnarray}
f_{p}(\tau) & = & \int^{0}_{-\infty}dzg_{p}(z)P(\tau,z)  \\   \nonumber
     		& =  &  \left(\frac{2}{3}\pi\tau^{3}\right)^{-1/2}
\exp\left(-\frac{3}{2\tau}\right).
\end{eqnarray}
The solid at the polar-cap surface consists of linear molecular chains of ions, 
parallel with ${\bf B}$ and normal to the surface, probably ordered so as to form a 
three-dimensional lattice. 
An approximate upper limit for $D$ can be obtained from the proton thermal kinetic 
energy and from the ion separation $a_{s} \approx 5\times 
10^{-9}Z^{1/2}_{26}B^{-1/2}_{12}$ cm found from the calculations of Medin \& Lai 
(2006).  It is,
\begin{eqnarray}
D = a_{s}\sqrt{\frac{1}{\beta m_{p}}} \approx 4\times 
10^{-2}\sqrt{\frac{T_{6}Z_{26}}{B_{12}}} \hspace{3mm}{\rm cm}^{2}{\rm s}^{-1},
\end{eqnarray}
where $\beta^{-1} = k_{B}T$ and $T_{6}$ is the polar-cap surface temperature in 
units of $10^{6}$ K.  This is based on the reasonable assumption that, within the 
solid structure, the proton potential barriers are at most $\sim k_{B}T$.  For 
$z_{p} = 10 l_{r}$ and $T_{6} = 1$, the characteristic time for diffusion to the 
surface is then $t_{p} \approx 10^{-5}$ with the values of $l_{r}$ given in Table 
1.

The condition of the surface at such temperatures is that it remains solid. The 
melting temperatures predicted by the one-component Coulomb-lattice condition given 
by Slattery, Doolen \& DeWitt (1980) are close to $10^{7}$ K at $B = B_{c}$ and are 
consistent with the melting temperature shown in Fig. 1 of the paper by Potekhin et 
al (2003). (We shall see in Section 4 that the longitudinal thermal conductivity is 
so large that the same statement can be made with confidence of matter at the 
shower maximum $z = z_{p}$.) Protons reaching the surface are accelerated parallel 
with the magnetic flux density ${\bf B}$ either immediately or, if their flux 
exceeds the corotation value, from the top of the gravitationally-bound proton 
atmosphere that is then formed.

It is also worth noting that the evaporation of ions at the surface is determined 
purely by the thermal excitation of lattice degrees of freedom.  Direct interaction 
with the inward flux of photons produced by curvature radiation from reverse 
electrons is not significant in this respect.  Photo-disintegration of nuclei 
within a small number of lattice planes immediately below the surface has 
negligible probability.  Photoelectric or bound-free cross-sections for the removal 
of electrons from ions are large, but the momentum transfer to the nucleus is 
determined by the Fourier transform of the bound electron wave function and its 
maximum value is therefore of the order of $137\hbar a^{-1}_{0}\sqrt{B/B_{c}}$, 
where $a_{0}$ is the Bohr radius.
The equivalent recoil energy is $140A^{-1}BB^{-1}_{c}$ eV for an ion of mass number 
$A$, which is negligible.

The magnitude of the photoelectric cross section is itself a source of reverse 
electron flux (see Jones 1981).  Electrons removed from partially-ionized atoms 
accelerated through the polar-cap blackbody radiation field form a significant flux 
at the polar-cap surface. Given the large values of $W_{p}$, a direct consequence 
is that the time-averages of the electron-positron, ion and proton current 
densities  $J_{e}$, $J_{Z}$ and $J_{p}$ on open magnetic flux lines  satisfy $J_{p} 
\gg J_{e,Z}$.  The time-averaged particle flux above the polar cap consists almost 
entirely of protons.

\subsection{The structure of the proton and ion atmosphere}

Ion cohesive energies are by no means negligible at $B\approx B_{c}$. Those 
calculated by Medin \& Lai (2006)  for $Z = 26$ can be expressed approximately as 
$E_{c} = 0.16B^{0.7}_{12}$ keV in the interval $10 < B_{12} < 100$ which is of 
interest here, and at the lower end of this interval are in fair agreement with the 
cohesive energy of $0.92$ keV for $Z = 26$ at $B_{12} = 10$ calculated by Jones 
(1985) using a different model for the three-dimensional lattice.  Thus $\beta 
E_{c} \gg 1$ and, in consequence, there is a large density discontinuity at the 
transition between the solid phase and the very thin gravitationally-bound 
atmosphere.  At lower magnetic fields, such that $\beta E_{c} \sim 1$, the 
discontinuity is less obvious and the atmosphere more dense.

The part of this atmosphere that is in local thermodynamic equilibrium exists  
within $0 < z < z_{1}$, where we assume that $z_{1}$ is the surface of last 
scattering, roughly defined here as the altitude beyond which number densities are 
so small that scattering of electrons and ions can be neglected.  The proton number 
density in this interval is in general many orders of magnitude smaller than those 
of ions and electrons which therefore determine the atmospheric structure.  Let us 
assume for the purposes of illustration that there is present only one ion species 
and that it is partially ionized with charge $\tilde{Z}$ and mass number $A$.  The 
standard relations between number densities and chemical potentials for free 
Boltzmann particles (allowing for non-zero $B$ in the case of electrons), with the 
constraint of electrical neutrality (strictly charge density $\rho_{0}$), require 
that in local thermodynamic equilibrium there must be a very small electrostatic 
potential $\Phi(z)$ given by
\begin{eqnarray}
e\Phi(z) = - \frac{1}{\tilde{Z} + 1}\left(M_{A,\tilde{Z}} - m\right)gz,
\end{eqnarray}
where $g$ is the local gravitational acceleration.  This result is not much changed 
if allowance is made for the presence of different ionization states (see Jones 
1986).  The scale height of the atmosphere is very small and, in the general case, 
the structure is of successive layers of increasing $\tilde{Z}/A$ as the altitude 
$z$ increases.  Equation (25) can be used to obtain the scale height of the very 
small proton component.  It is negative, showing that protons cannot be in 
thermodynamic equilibrium at $z < z_{1}$ and are accelerated across the surface of 
last scattering.

The proton number density equivalent to the corotational charge density is given by 
$N_{p}ev_{p}/c = \rho_{0}$ and, even for polar-cap thermal values of velocity 
$v_{p}$, is much smaller than ion densities at $z_{1}$.  Thus as particles move out 
in Landau states parallel with ${\bf B}$ to $z > z_{1}$, the condition of 
electrical neutrality again requires an electrostatic potential that is independent 
of the small proton number and is also given precisely by equation (25).  The 
proton potential energy at $z > z_{1}$ is
\begin{eqnarray}
m_{p}(z - z_{1})g + e\Phi(z) - e\Phi(z_{1}) =    \nonumber  \\
m_{p}(z - z_{1})g - \frac
{1}{\tilde{Z} + 1}\left(M_{A,\tilde{Z}} - m\right)(z - z_{1})g,
\end{eqnarray}
which is evidently negative. The relatively small numbers of protons are again 
accelerated whilst ions, with smaller charge-to-mass ratios, are retained by the 
gravitational field. Thus the protons are preferentially selected for the process 
of inertial acceleration that follows.  If the flux of protons at $z = 0$ is 
greater than is needed for the corotational current density, the surplus forms an 
atmosphere at $z > z_{1}$.  This ordering has significant consequences for plasma 
formation, as will be seen in Section 4.

\subsection{Reverse electron flux from photoelectric ionization}

Inertial acceleration under the ${\bf E}_{\parallel} = 0$ polar-cap surface 
boundary condition was first described by Michel (1974).  Owing to current 
conservation, an excess of positive charge is present immediately above the surface 
in the non-relativistic stage of ion acceleration.  This is the source of an 
electric field,
\begin{eqnarray}
E_{i} \approx \left(\frac{8\pi\rho_{0}Mc^{2}}{\tilde{Z}e}\right)^{1/2}
  = 5.0\times 10^{4}\left(\frac{-AB_{12}\cos\psi}{P\tilde{Z}}\right)^{1/2}
\end{eqnarray}
expressed in electrostatic units (esu) at $z > z_{i}$, which is the roughly defined 
altitude at which ion motion becomes fully relativistic,
\begin{eqnarray}
z_{i} \approx \left(\frac{Mc^{2}}{18\pi\rho_{0}\tilde{Z}e}\right)^{1/2}
 = 4.1\times 10^{1}\left(\frac{-AP}{B_{12}\tilde{Z}\cos\psi}\right)^{1/2}
  {\rm cm}
\end{eqnarray}
where $P$ is the rotation period and $\psi$ is the angle between ${\bf \Omega}$ and 
${\bf B}$.  (In obtaining these approximate results, the ion charge $\tilde{Z}$ has 
been assumed constant in the interval $0 < z < z_{i}$.)

The change of ionization from partial ($\tilde{Z}$) to complete ($Z$) is the result 
of photoelectric absorption of blackbody photons whose energies have been boosted 
by Lorentz transformation to the rest frame of the accelerated ion.  A simple model 
of the process has been described previously in which
transitions occur sequentially in order of increasing electron separation energy at 
altitudes for which unit optical depth of blackbody radiation field for a 
particular transition is reached.
In this way, a compact approximate expression for the mean reverse electron energy 
at the polar-cap surface per accelerated unit positive charge can be obtained 
(Jones 1981; see equations 15 - 22, and Table 2 of that paper). The
prefactor in the expression for the electron
separation energies assumed there was obtained for smaller magnetic fields ($B_{12} 
\sim 1$) than those considered here, but comparison with the recent calculations of 
Medin \& Lai (2006) for multiply ionized C and Fe ions shows that 
 separation energies at $B > B_{c}$ can found with sufficient accuracy for present 
purposes by scaling from that expression. Following this procedure, the mean 
electron energy per accelerated unit nuclear charge is estimated to be
\begin{eqnarray}
\left(\frac{Z - \tilde{Z}}{Z}\right)\bar{\epsilon}_{e} \approx
 5.6\times 10^{4}Z^{0.85}_{26}(0)B^{0.5}_{12}T^{-1.0}_{6} \hspace{3mm} mc^{2},
\end{eqnarray}
in which $Z(0)$ is the mean nuclear charge at the surface $z = 0$.  These values 
assume the acceleration field is uniform at altitudes small compared with the 
polar-cap radius, but are quite insensitive to variations in its strength because 
photoionization occurs quite promptly when the blackbody radiation field is 
boosted, in the rest frame of the accelerated ion, to the electron separation 
energy.  The altitude at which this occurs has little effect on the
reverse-electron energy flux.  We assume a value $E_{\parallel} = E_{i} = 10^{6}$ 
esu.
Equation (29) obviously fails for very light ions, $Z < 4-5$, which are likely to 
be completely ionized in the LTE region at $z < z_{1}$.
It can be combined with an approximate parameterization of the values for $W_{p}$ 
given here in Table 1,
\begin{eqnarray}
W_{p} = 7.0\times 10^{-5}\langle Z^{-0.76}_{26}\rangle_{sm}B^{0.12}_{12} 
\hspace{3mm} (mc^{2})^{-1},
\end{eqnarray}
where square brackets denote the average nuclear charge in the region of the shower 
maximum,
to obtain an expression for the number of protons produced per unit positive 
nuclear charge accelerated.  It is,
\begin{eqnarray}
K = K_{0}Z^{0.85}(0)\int g_{p}(z)dz    \nonumber
\end{eqnarray}
in which,
\begin{eqnarray}
K_{0} = 0.24\langle Z^{-0.76}_{26}\rangle_{sm}B^{0.62}_{12}T^{-1.0}_{6}.
\end{eqnarray}
We note again that this is invalid for very light ions, $Z < 4-5$.  It is also not 
valid at much smaller fields where $K$ tends to an asymptotic value.  Here, $W_{p}$ 
is given by the zero-field value of the shower parameter $y$ (see Section 2.5) and 
by placing $B_{12} = 1$, so that equation (30) is replaced by
$W_{p} = 3.8\times 10^{-5}\langle Z^{-0.76}_{26}\rangle_{sm} (mc^{2})^{-1}$.  
Electronic separation energies concerned here approach asymptotically their 
zero-field values at $B_{12} < 1$ so that equation (29) should be evaluated at 
$B_{12} = 1$ to obtain an estimate of the asymptotic value of $K_{0}$,
\begin{eqnarray}
 K_{0} = K_{0as} = 0.13\langle Z^{-0.76}_{26}\rangle_{sm}T^{-1.0}_{6}.
\end{eqnarray}

Equations (31) and (32) are an under-estimate of $K$ owing to our neglect here of a 
further source of reverse electrons that is difficult to quantify, specifically 
pair creation following inverse Compton scattering of blackbody photons by the 
photoelectrons or the consequent conversion of curvature radiation photons at 
higher altitudes.  The electron and photon momenta are so oriented that the 
necessary electron Lorentz factor, $10^{4}-10^{5}$, is easily reached.  For the 
significance of the outward accelerated positrons as a source of pairs, we refer to 
Harding \& Muslimov (2002).

\section[]{Application to polar caps}

Neutron stars with spin direction such that ${\bf \Omega}\cdot{\bf B} < 0$ and 
polar-cap surface boundary condition ${\bf E}_{\parallel}$ = 0 must
presumably exist but the extent to which they are present in the observed 
population of radio pulsars is an unsolved problem.  The physical existence of case 
(iii) of Section 1, the boundary condition ${\bf E}_{\parallel} \neq 0$, is less 
certain in the observed neutron star population
 and is a question to which proton production in polar-cap reverse electron flow is 
very relevant.  This section attempts to estimate the interval of polar-cap 
magnetic field that would support this condition and to investigate the nature of 
the plasma formed.  It does not attempt to consider the actual radio emission 
process.

The distribution of the inferred polar-cap dipole field strengths for the observed 
radio pulsar population contained in the ATNF Catalogue (Manchester et al 2005) 
indicates that
a large fraction are likely to be too small to support case (iii).  But it is 
interesting to note, though with reservations about the possibility of unknown 
selection bias, that the nulling radio pulsars listed in Tables 1 and 2 of the 
paper by Wang, Manchester \& Johnston (2007) have a distribution significantly 
displaced toward larger fields in comparison with the complete ATNF catalogue 
population.
The distribution of $\log_{10}B$, where $B$ is here the listed polar-cap field, for 
the whole population of $1486$ radio pulsars is slightly skewed toward low values 
but has a mean of $11.98$ and a median of $12.06$. With neglect of the low-$B$ 
tail, the distribution has
a standard deviation of approximately $0.50$.  The $63$ pulsars for which Wang et 
al give a finite nulling fraction have a mean of $12.41$ and a median of $12.45$.            
The expected standard deviation of the mean for a randomly selected set of $63$ 
pulsars is approximately $0.07$ which shows that the set of nulling pulsars have 
field strengths that are significantly larger than the general population.

We shall see that if the polar-cap fields were identical with their inferred dipole 
values, few if any pulsars would be able to support the boundary condition ${\bf 
E}_{\parallel} \neq 0$, given the ion cohesive energies referred to at the end of 
Section 3.2.  But in this argument, it is implicit that the dipole origin is 
positioned at the centre of the star.  Outward displacement of the origin along the 
dipole axis toward the surface of the star would produce considerable amplification 
of the field at one polar cap and diminution at the other without changing 
significantly the dipole field at the light cylinder. Beyond this simple case there 
are, of course, many other possible configurations with enhanced polar-cap fields.

The proton formation rates given in Table 1 are high and any atmosphere formed 
under boundary condition (ii) is fractionated in the manner described in  Section 
3.2. Thus the protons undergo inertial acceleration first and, if exhausted, are 
followed by ions in order of decreasing $\tilde{Z}/A$ if they are present in the 
atmosphere.  Because the time-averaged total open magnetic flux-line current 
density consists almost entirely of protons and must satisfy the condition $J < 
\rho_{0}c$ it follows that the values of $W_{p}$ lead directly to an upper limit 
for the reverse electron-photon energy flux at the polar-cap surface.  The 
temperature limit derived from this for a typical value $W_{p}mc^{2} = 2\times 
10^{-4}$ is,
\begin{eqnarray}   
T_{max}= 6.2\times 10^{5}\left(\frac{-B_{12}\cos\psi}{P}\right)^{1/4} 
\hspace{2mm}{\rm K}.
\end{eqnarray}
Because it has been obtained from the time-averaged energy flux, it can in 
principle be exceeded in a time-variable state.  But in a time-independent state it 
can be compared with the critical temperature $T_{c}$ above which the ion thermal 
emission rate is high enough to maintain the case (ii) boundary condition ${\bf 
E}_{\parallel} = 0$.  In terms of the work function, this is approximately 
$k_{B}T_{c} = 0.030E_{c}$.  The values of Medin \& Lai (2006) for $Z = 26$ and $10 
< B_{12} < 100$ can be expressed as $E_{c} = 0.16B^{0.7}_{12}$ keV and
give $T_{c} = 5.6 \times 10^{4}B_{12}^{0.7}$ K.  This can be compared with 
$T_{max}$ or with $T_{res}$, the polar-cap blackbody temperature in the absence of 
reverse-electron flux. (This is approximately the whole-surface blackbody 
temperature corrected to the local proper frame.)  We can see that, regardless of 
the reverse electron flux, the
${\bf E}_{\parallel} \neq 0$ boundary condition can be satisfied only in very old 
pulsars or in those where the polar cap field is of the order of $B_{c}$.

\subsection{Medium time-scale variability}

The observed field-strength distribution for nulling pulsars indicates that it may 
be worth considering whether or not
the variability time-scales associated with
nulling can be a feature of the boundary conditions assumed in case (iii). But 
first, in Section 4.1, we shall consider case (ii).  As we have seen in Section 3, 
photodisintegration reduces nuclear mass numbers in the region of the shower 
maximum.  Protons diffuse to the surface with negligible probability of nuclear 
interaction.  A fraction $\epsilon_{n}$ of neutrons also escape nuclear capture and 
reach the surface. Thus nuclei fixed in a lattice plane move gradually toward the 
surface with a small velocity $v$.
(Our reference to a lattice is merely a convenience.  There will be order parallel 
with ${\bf B}$, but also a considerable density of point defects produced by 
neutron scattering.)  Instability against the growth of curvature radiation pairs 
requires a minimum $E_{\parallel}$ and hence, broadly, that the current density $J$ 
should not exceed a critical value $J_{crit}$.
We consider first case (ii) with current density $J > J_{crit}$ so that there is no 
instability against pair formation by curvature radiation.  Solution of equation 
(1), with time-independent boundary conditions and inclusion of inertial 
acceleration, determines $J$ which is constant.  But intuition suggests that the 
distribution of nuclear charge with depth is not necessarily time-independent 
because the surface charge $Z(0,t)$ is determined by the surface state at a 
previous time $t - t_{sm}$, where $t_{sm}$ is the time interval in which nuclei 
move from the shower maximum to the surface.
Let us assume, for simplicity, that $\epsilon_{n} = 1$. 

It will be convenient in this Section to describe the motion of nuclei toward the 
surface in terms of a Lagrangian depth $\tilde{z}$ which moves with velocity 
$v(z,t)$ relative to $z$ so that $\tilde{z} = z$ at $t = 0$. The rate of change of 
nuclear charge within an element at this depth containing a fixed number of nuclei, 
$N\delta\tilde{z}$ where $N$ is given by equation (4), is
\begin{eqnarray}
N\delta\tilde{z}\frac{\partial Z}{\partial t^{\prime}} = 
 -g_{p}(z)\delta zK_{0}N(0,t^{\prime})Z^{1 + \nu}(0,t^{\prime})v(0,t^{\prime}),
\end{eqnarray}
where $\nu = 0.85$ from equations (31) and (32).
But $N \propto Z^{\alpha}$ so that the evolution of $Z$ satisfies,
\begin{eqnarray}
Z^{\alpha + 1}(-\infty) - Z^{\alpha + 1}(0,t) =   \hspace{2cm}  \nonumber \\
(\alpha + 1)
 \int^{t}_{-\infty}dt^{\prime}\delta(t - t^{\prime} - t_{sm})K_{0}Z^{\nu + \alpha + 
1}(0,t^{\prime}).
\end{eqnarray}
This expression contains the approximation $g_{p} = \delta(z - z_{p}) \approx 
v^{-1}(0,t^\prime)\delta(t - t^{\prime} - t_{sm})$ which is satisfactory provided 
$g_{p}$ is a sharply-peaked function displaced below the surface as is expected 
owing to the nature of shower development and to the LPM effect.  It is also 
unimportant that $t_{sm}$ is to some extent a function of $Z$-values in the 
interval $z_{p} < z < 0$.
The initial value of $Z$ is $Z(-\infty)$ and it is assumed constant.  A 
time-independent solution  of equation (35) exists, given by,
\begin{eqnarray}
\left(\frac{Z(0)}{Z(-\infty)}\right)^{1 + \alpha} = \frac{1}{1 + (1+\alpha)K}
  \nonumber
\end{eqnarray}
But assuming  $Z(-\infty) = 26$, and a time-independent charge at the surface so 
small, $Z(0) = 5$, that equations (29) and (31) begin to fail because few 
accelerated ions have any bound electrons, we find a self-consistent value $K = 
5.5$ for $\alpha = -0.7$.
However, this state is not stable because for a charge fluctuation $\delta Z(0,t)$, 
equation (35) becomes a homogeneous Volterra equation of the second kind with no 
non-zero square-integrable solution. The charge fluctuation satisfies,
\begin{eqnarray}
\delta Z(0,t) = - (\nu + \alpha + 1)K_{0}Z^{\nu}(0,t - t_{sm})\delta Z(0,t - 
t_{sm}),
\end{eqnarray}
and for the above value of $K$, grows indefinitely with alternating sign. 
Ultimately, equation (36) fails and the system alternates between intervals of high 
nuclear charge at the surface and intervals in which the surface nuclear charge is 
so low that the reverse electron flux is negligible.  The basic unit of time in 
which $t_{sm}$ is measured is the time $t_{rl}$, defined in Section 3, for the 
removal of one radiation length of matter at the corotational current density. This 
is certainly within the orders of magnitude associated with nulling intervals.  The 
behaviour in time of a complete polar cap is, of course, much more complex because 
the interval of depth $z_{p}$ considered here is extremely small compared with the 
cap radius $u_{0}$, and the question arises of the extent (if any) of correlation 
between different areas of the cap. 

It is obvious that there must be variability with time-scales of the order of 
$t_{rl}$ in case (iii).  Although there is little information about the 
$Z$-dependence of the ion work function $E_{c}$, we can be confident that it must 
ultimately decrease to a very small value in the limit $Z \rightarrow 1$. 
Thus there exists a critical value of $Z$ at which the boundary condition ${\bf 
E}_{\parallel}\neq 0$ fails locally in some region of the polar cap and this must 
ultimately be reached as a consequence of proton emission. But the acceleration 
field is a function
of the condition of the whole polar cap and the overall level of complexity is such 
that we make no attempt here to proceed further with this analysis except to note 
that time-independent case (iii) solutions are not possible.

\subsection{Factors relevant to short time-scale variability}

There are several characteristic times that are important factors in examining the 
possibility of short time-scale variability of plasma production in cases (ii) and 
(iii).
The first is the proton diffusion time $t_{p}$ derived from $z_{p}$ and $D$.  Then 
the initial e-folding growth time for a curvature-radiation electron-positron 
cascade must be of the order of $t_{ee} = 2l_{a}/c$, where $l_{a}$ is the open- 
magnetosphere acceleration length interval concerned.  The third parameter is 
$t_{cond}$ the time-scale for thermal diffusion from the shower maximum $z_{p}$ to 
the surface.

For an estimate of $t_{cond}$, we require the surface temperature fluctuation from 
$T$ to  $T + \delta T(0,t)$ caused by a fluctuation $\delta 
X(z^{\prime},t^{\prime})$ in the thermal power input at depth $z^{\prime} < 0$ 
below the polar-cap surface at time $t^{\prime}$.  The correct Green function for 
this particular problem is cumbersome (see Carslaw \& Jaeger 1959), but the 
characteristic value of $t - t^{\prime}$ is, for present purposes, satisfactorily 
approximated by $t_{cond} = Cz^{2}_{p}/2\lambda_{\parallel}$.  Here 
$\lambda_{\parallel}$ is the longitudinal coefficient of thermal conductivity  and 
$C$ is the specific heat of surface matter, for which an upper limit given by the 
Dulong and Petit law is assumed.  Transport coefficients at $B > B_{c}$ in 
neutron-star envelopes and crusts have been calculated by Potekhin (1999). The 
longitudinal coefficient of thermal conductivity is not a particularly rapidly 
varying function of $B$ and $T$ and, at $Z = 26$, $B = 3B_{c}$ and $T = 10^{6}$ K, 
is approximately $6 \times 10^{14}$ erg cm$^{-1}$ s$^{-1}$ K$^{-1}$. With the ion 
number density given by equation (4) and a shower depth of $z_{p} = 10 l_{r}$, the 
time is $t_{cond} \approx 5 \times 10^{-10}$ s, which is short compared with 
$t_{ee}$ and with the estimates of $t_{p}$ obtained from equations (21) - (24).  It 
is also possible to confirm that, for the expected power inputs associated with the 
derived temperature limit given by equation (33), the temperature gradient between 
the shower maximum and the surface is no more than of the order of $10^{6}$ K 
cm$^{-1}$, which is negligible.

Thus in relation to variability on time-scales set by $t_{ee}$ or $t_{p}$, we can 
assume that the surface temperature fluctuations are instantaneous functions of the 
reverse-electron power input to the polar-cap surface.  This is one of three 
important factors.  The second is the mass-number fractionation of any atmosphere 
which was described in Section 3.2.   The final factor is that the accelerated 
protons produce a quite negligible reverse-electron flux, as can be seen by 
examining their electromagnetic interactions with the thermal radiation field. 
Protons accelerated to energies  of the order of $10^{3}$ GeV are well below the 
threshold for pair-creation, directly or by Compton scattering, and in any case the 
cross-sections concerned are small.  However, accelerated ions are also below the 
pair-creation threshold but, as we have seen in Section 3.3, those with $Z$ 
sufficiently high to have bound electrons produce a reverse-electron flux by a 
sequence of photo-dissociation reactions with blackbody photons which increase the 
ion charge from $\tilde{Z}$ toward $Z$
(Jones 1981; see Medin \& Lai 2006 for atomic ionization energies at $B \approx 
B_{c}$).

Possible sources of the electron-positron current density $J_{e}$ in neutron stars 
with ${\bf \Omega}\cdot{\bf B} < 0$ need consideration and we refer to the recent 
review of Harding \& Lai (2006).  The growth of curvature radiation cascades 
requires a large $E_{\parallel}$, that is, a partial vacuum above the polar cap ($J 
< J_{crit}$), with the possibility that $J_{crit} = 0$.  A further  source is 
inverse Compton scattering of the blackbody radiation field by the reverse electron 
flux from accelerated ions, followed by magnetic conversion of the scattered 
photons. This was mentioned briefly at the end of Section 3.3 and we refer to 
Harding \& Muslimov (2002) for further details.

\subsection{Short time-scale variability; $E_{\parallel} = 0$}

A first investigation of short time-scale variability in case (ii) with $T_{max} > 
T_{c}$ is most easily made, as in Section 4.1, with the assumption of a 
time-independent $J$, with negligible $J_{e}$. The proton current is then,
\begin{eqnarray}
J_{p}(t) = K \int^{t}_{-\infty}dt^{\prime} f_{p}(t - t^{\prime})
\left(J - J_{p}(t^{\prime})\right),
\end{eqnarray}
where here the parameter $K$ is a constant given by equations (31) or (32). The 
time $t$ in this Section is dimensionless and in units of $t_{p}$.
Equation (37), with $J$ determined by solution of equation (1) for the immutable 
boundary condition $\Phi = 0$ on the surface separating open from closed magnetic 
flux lines, may not be a completely adequate description of short time-scale 
variability of the system, but is a useful starting point. It has the obvious 
time-independent solution $J_{p} =KJ/(K+1)$ but 
it can be seen that for any small fluctuation in proton current density, equation 
(37) reduces to a homogeneous Volterra equation of the second kind having no 
square-integrable non-zero solution.  For the present problem as opposed to that of 
Section 4.1, we choose to introduce a short time-scale fluctuation as an 
inhomogeneous term $\delta J_{p0}(t)$ centred on $t = 0$, and rewrite the equation 
as,
\begin{eqnarray}
J_{p}(t) = KJ +\delta J_{p0}(t) - K\int^{\infty}_{0}d\tau f_{p}(\tau)
J_{p}(t-\tau),
\end{eqnarray}
with $\tau = t - t^{\prime}$.  We can see that solutions that are not 
square-integrable exist for $K > 1$.  For example, the case $f_{p}(\tau) = 
\delta(\tau - \tau_{0})$ gives the obvious iterative solution,
\begin{eqnarray}
J_{p}(t) = \frac{K}{K+1}J +\delta J_{p0}(t) - K\delta J_{p0}(t-\tau_{0}) +
\\   \nonumber
\hspace{3cm}   K^{2}\delta J_{p0}(t - 2\tau_{0}) -  . . . .
\end{eqnarray}
More generally, we can introduce the Fourier transform of $J_{p0}$ into the formal 
iterative solution of equation (38) and sum to infinity to obtain,
\begin{eqnarray}
J_{p}(t) = \frac{K}{K+1}J + \frac{1}{2\pi}\int^{\infty}_{-\infty}d\omega
\frac{\delta J_{p0}(\omega)\exp(-i\omega t)}{1 + Kf_{p}(\omega)},
\end{eqnarray}
where,
\begin{eqnarray}
f_{p}(\omega) = \int^{\infty}_{0}d\tau f_{p}(\tau)\exp(i\omega\tau).
\end{eqnarray}
As expected, the trivial case of equation (38) can be recovered by summation over 
the residues of those poles in the integrand that are in the upper half of the 
complex-$\omega$ plane.  On the other hand, there exist simple functions, such as 
$f_{p}(\tau) = a\exp(-a\tau)$ with constant $a$ that give no such poles.  This 
indicates that the behaviour of $f_{p}(\tau)$ at small $\tau > 0$ is crucial, so 
that for non-trivial functions, such as equation (23), there appears to be no 
alternative to a numerical search for zeros in the denominator.  Thus for $f_{p}$ 
given by equation (23) we find, for $K = 100$,
poles at $\omega = \pm 4.85 + 1.88i$. For smaller $K$ they approach, but do not 
reach, the real axis; for $K = 25$, the poles are at $\omega = \pm 3.4 + 0.1i$.

Our conclusion is that there is no doubt that the possibility of instability exists 
for large $K$ with exponential growth of short time-scale fluctuations until the 
condition $0 < J_{p} < J$ is no longer satisfied and equation (37) fails.  But 
unfortunately, certainty is not possible because we do not have sufficient 
knowledge of the actual shape of the formation depth distribution $g_{p}$ at depths 
$z_{p} < z < 0$ but close to the surface.  The existence of poles in the  upper 
half of the complex-$\omega$ plane in equation (40) appears to depend crucially on 
the time interval between the creation of protons and their arrival at the surface.  
This almost certainly depends on the LPM effect considered in Section 2.4.

The behaviour of the system when equation (37) fails is easy to see. Suppose, for 
example that $T_{res} > T_{c}$, as must be the case in many pulsars, so that the 
boundary condition ${\bf E}_{\parallel} = 0$ is always maintained, also that 
$J_{crit} < J$ so that growth of curvature radiation is not possible.  Instability 
resulting in excess proton production leads to the formation of a proton atmosphere 
at $z > z_{1}$, as described in Section 3.2, and the surplus 
results in a current density containing only the proton component of magnitude  
$J_{p} = J$. For the interval of time within which this state persists there is no 
reverse-electron flux, no further proton formation, and the surface temperature 
falls promptly to $T_{res}$ in a time no more than an order of magnitude greater 
than $t_{cond}$.  At the end of this interval, the proton atmosphere is exhausted 
and $J_{p}$ falls abruptly to a value given by the tail of the distribution of 
$f_{p}$, much below its time-independent value, $KJ/(K+1)$.  Consequently, there 
will be a prompt burst of ion acceleration and of outward moving positrons produced 
by the inverse Compton scattering process (see Harding \& Muslimov 2002) unless 
$T_{res} < T_{c}$.  Owing to the time-delay, this can be expected to lead to the 
formation of a further proton excess.  Strict periodicity is not to be expected 
because, as in the case of medium time-scale variability, there is no obvious 
reason why the state of a very thin layer, of height $z\sim z_{1}$ should be 
uniform over the whole polar-cap area.

\subsection{Short time-scale variability; ${\bf E}_{\parallel} \neq 0$}

In a time-independent state, this boundary condition requires $T_{max} < T_{c}$, 
where $T_{max}$ is the polar-cap blackbody temperature given by equation (33).  But 
in a state of short time-scale variability, it can be satisfied intermittently 
provided the less restrictive condition $T_{res} < T_{c}$ is valid.  We can begin 
by considering a steady state with $J > J_{crit}$ as in the previous section, with 
polar-cap temperature $T > T_{c}$.  The instability leads to a state of excess 
proton production and to a very prompt decrease in temperature to $T_{res}$ in a 
time probably one or two orders of magnitude larger than $t_{cond}$ but still much 
smaller than $t_{p}$
and to a similarly fast partial collapse of the ion LTE atmosphere, so that only 
protons and any low-$Z$ ions remain.  When these are exhausted, the current density 
$J$ promptly decreases toward zero and if $J_{crit}$ has no finite value (the star 
cannot support pair creation by curvature radiation at any current density) it 
would appear that plasma generation ceases.   But a finite $J_{crit}$ allows pair 
creation when $J < J_{crit}$.  Thus $J_{e} + J_{Z}$ increases rapidly and a further 
proton excess is generated.

The conclusion to be drawn here is that, given short time-scale variability, the 
important polar-cap surface temperature is $T_{res}$ which should be compared with 
the work-function related temperature $T_{c} = 5.6\times 10^{4}B^{0.7}_{12}$ K.  
Thus for $T_{res} = 5\times 10^{5}$ K, a surface magnetic flux density $B_{12} > 
23$ is required for the boundary condition ${\bf E}_{\parallel} \neq 0$ to be 
satisfied.

\section[]{Conclusions}

There seems to be no known reason why neutron stars with ${\bf \Omega}\cdot{\bf B} 
< 0$ and hence positive polar-cap corotational charge density should not be formed  
at a rate of the same order as those with the opposite sign. The purpose of this 
paper has been to investigate some of the features of plasma formation at the polar 
cap to see if these correlate with the properties of any subset of the observed 
population of radio pulsars.  Specifically, we have attempted to find the formation 
rate for protons in the electromagnetic showers produced by the reverse electron 
flux at the polar cap.  These protons diffuse rapidly to the surface and, having 
both a negligibly small work function and the highest charge to mass ratio, are 
preferentially accelerated upward by the  electric field component parallel with 
${\bf B}$.  Proton formation is necessarily accompanied by evolution of the nuclear 
charge within the very thin surface layer in which the showers occur.

The proton formation rate was estimated previously (Jones 1981) at magnetic flux 
densities of $B_{12} \sim 1$ but the present paper has been addressed to the 
problem at much higher fields ($B \sim B_{c}$) owing to the realization that such 
values, though infrequent, are not unknown in the distribution of inferred 
dipole-field strengths.  Additionally, an important subset, those radio pulsars 
exhibiting the nulling phenomenon, have higher than average fields, as noted at the 
beginning of Section 4.  Values of the parameter $K$, defined as the number of 
protons formed per unit positive accelerated nuclear charge, are given in Table 1 
for $B \sim B_{c}$, but its parametrization in equation (31) is not valid at small 
values of $B$ for which its asymptotic value $K_{as}$ is given by equation (32).

Proton formation and the evolution of nuclear charge have the consequence that 
particle fluxes under the case (ii) boundary condition defined in Section 1 display 
much  more complexity than might have been supposed.  In particular, there is an 
instability, given by equation (36), which is realized in almost any radio pulsar, 
giving time variability typically over intervals of the order of $t_{rl}$.  This is 
the time in which one radiation length of matter flows from the polar cap and is 
within the orders of magnitude associated with nulling intervals.   Owing to the 
proton formation-depth distribution, fluctuations in the surface nuclear charge 
grow until the system alternates between having high values close to $Z(-\infty)$ 
and  nuclear charges so low that the reverse electron flux is negligible.
The latter state has no source of electron-positron pair creation at the polar cap 
and may possibly be associated with the intervals of null emission. But the state 
of the whole polar cap is almost certainly much more complex than a simple 
transition between high and low surface nuclear charge.  The total depth of the 
shower formation region and of the LTE atmosphere is very many orders of magnitude 
smaller than the polar cap radius $u_{0} \sim 10^{4}$ cm.  The question then is why 
the time-variable state of the surface and atmosphere should be in phase over the 
whole polar-cap area.  No serious attempt has been made here to investigate this 
problem but there seems to be no reason to expect such phase uniformity. Chaotic 
behaviour appears more probable.
Similar time-variability and complexity are expected in case (iii).

Instability leading to short time-scale variability is also possible. The relevant 
unit of time is $t_{p}$, the characteristic time for proton diffusion from depth 
$z_{p}$ to the surface.  Here, growth of the instability results in alternation 
between two states of which the first is an accelerated proton current density 
$J_{p} = J$ with $J_{e} = J_{Z} = 0$, no reverse electron energy flux and a 
polar-cap surface temperature equal to $T_{res}$.  Abrupt exhaustion of the proton 
atmosphere then allows rapid growth in $J_{e}$ and $J_{Z}$ and the formation of a 
short pulse of electron-positron pair formation.  This instabilty appears to be 
common to both cases (ii) and (iii).  It is also interesting because it defines the 
temperature at which the case (iii) boundary condition ${\bf E}_{\parallel} \neq 0$ 
must be satisfied.  This must be $T_{res}$ and not the temperature $T_{max}$ of 
equation (33) defined by the time-averaged maximum reverse-electron energy flux.  
The condition is then $T_{res} < T_{c}$.  From our parametrization of the work 
functions found by Medin \& Lai (2006), the critical temperature is $T_{c} = 
5.6\times 10^{4}B^{0.7}_{12}$ K, and we find that the required polar-cap surface 
magnetic flux density is $B_{12} > 2.3T^{1.4}_{res5}$ in which $T_{res5}$ is the 
polar-cap temperature, in units of $10^{5}$ K, in the absence of a reverse-electron 
energy flux.  This unit of $T$ is not inappropriate for the whole-surface 
temperatures of the older radio pulsars and the result shows that the case (iii) 
boundary condition can be realized at much lower magnetic flux densities than was 
previously thought.  At the same time, the similarities between the instabilities 
found in cases (ii) and (iii) suggests that, in terms of observable phenomena, the 
distinction between the two boundary conditions may not be of great significance.  
It is to be emphasized that, in both cases, the cessation of ion flow is not 
temperature-controlled (by the ionic analogue of Richardson's formula).  Rather, it 
is controlled by excess proton production and by the consequent formation of a 
proton atmosphere at $z > z_{1}$.
In case (iii), only the onset of ion flow is a result of surface heating.

An interesting state is that of case (ii) with the parameter $K < 1$ derived from 
the $B_{12} \ll 1$ expression for the asymptotic limit given by equation (32).  
This can be realized at polar-cap temperatures greater than $10^{6}$ K, 
particularly if this temperature is supported by a large value of $T_{res}$.  With 
reference to equation (36), we can see the associated proton and ion current 
densities are stable and time-independent.  Monoenergetic ion beams do interact 
with an electron-positron plasma (see Cheng \& Ruderman 1980) but we anticipate  
little if any electron-positron pair production by inverse Compton scattering at 
$B_{12} \ll 1$  
and there may be no observable consequences of such a system. This may be 
particularly relevant to young neutron stars or to any older neutron stars that for 
other reasons have very high values of $T_{res}$.

A further comment can be made about the reverse electron  energy flux from any 
outer gap that exists. It is hard to make quantitative estimates of this, but there 
is every possibility that the consequent rate of proton formation would completely 
inactivate polar-cap pair formation as in nulling intervals. 

The time-variabilities considered so far have been found assuming a 
time-independent boundary condition $\Phi = 0$ on the surface separating open from 
closed magnetic flux lines. But we have not addressed the problem of how, or if, 
this boundary condition is maintained.  The proton diffusion time $t_{p}$
is little more than an order of magnitude greater than the typical transit time of 
order $u_{0}c^{-1}$ and a correct treatment of polar-cap phenomena may require not 
just the solution of equation (1) but solution of the full set of Maxwell equations 
with retardation.  Thus we do not exclude the existence of further instabilities 
beyond those considered here.
This comment may also be relevant to case (i) in which the corotational charge 
density is negative.  The considerations of this paper have no application to this 
class of neutron star.  The electron work function is so small that there is no 
doubt the condition ${\bf E}_{\parallel} = 0$ is satisfied at all times on the 
polar cap, also that electrons are freely available to maintain the $\Phi = 0$ 
boundary condition on the surface separating open from closed magnetic flux lines. 

It would be of interest to know how boundary conditions (i), (ii) and (iii) are 
distributed within the observed population of radio pulsars and neutron stars and 
whether or not the short time-scale temporal variability shown here to be a feature 
of cases (ii) and (iii) contributes to the observed microstucture of some radio 
pulses.  Perhaps systematic measurements of individual pulse structures with 
$10^{-6} - 10^{-5}$ s resolution would provide some insight.

\bsp

\label{lastpage}

\end{document}